\documentclass[10pt]{article}
\usepackage[margin=1in]{geometry}
\usepackage{amsmath}
\usepackage{abstract}
\usepackage{amsfonts,color}
\usepackage{amssymb}
\usepackage{graphicx}
\usepackage{hyperref}
\begin{document}
\title{\bf{Amplitude death by delay induced position coupling in a system of two coupled Van der Pol Oscillators}}
\author
{ Satadal Datta\\
Harish-Chandra research Institute, HBNI, Chhatnag Road, Jhunsi, Allahabad-211019, INDIA\\
\date{}
satadaldatta1@gmail.com}
\maketitle
\begin{abstract}
We consider a system of two interacting identical Van der Pol Oscillators in a simple harmonic potential well. The position coupling term between the oscillators is such that there is a finite delay, i.e; each system takes a finite time to react to the other one. We investigate the amplitude death in the presence of such interaction and we find out the amplitude death regions in the corresponding parameter space. 
\end{abstract}
\section{Introduction}
Coupled oscillatory systems are ubiquitous in nature as well as this type of systems have been studied in different research areas., such as, in cell biology\cite{a}, in modeling pacemakers\cite{b}, in understanding nonlinear oscillatory chemical reactions like Belousov-Zhabotinsky reaction\cite{c}\cite{d}, in studying synchronization in Josephson series array\cite{e}, in studying coupled nonlinear semiconductor lasers\cite{f} etc. Non linear oscillators like Van der Pol oscillators, in the absence of coupling, have limit cycle\cite{g} but when coupled to a different oscillator may exhibit a Hopf Bifurcation depending on the coupling strength and other relevant parameters in the coupled system, i.e; amplitude of the limit cycle becomes zero. This phenomena is called Amplitude death. Amplitude death is present in several different coupled systems\cite{h}-\cite{k}. In a coupled system, in general, the response of one oscillator to the other one takes some finite time  because of finite speed of propagation of signal from one oscillator to the other one and that's why there is always a delay in the interaction term of the coupled system. This delay effect is also present in several different physical systems\cite{l}-\cite{n}. Amplitude death in such a system, having a delay in the coupling term, is studied in a model of coupled nonlinear oscillators\cite{o}. Amplitude death occurs also in the presence of distributed delay\cite{p}. Bifurcations and stability in a system of Van der Pol oscillators with position coupling and with a delay in the interaction term is analyzed\cite{q}. Effects of time delay in the velocity coupling term in a system of two Van der Pol oscillators is also studied\cite{r}. In a system of a  pair of Van der Pol oscillators with a delay in both position as well as velocity coupling and also in the presence of a self-connection term, there exists amplitude death in the parameter space; this can be found by stability analysis\cite{s}.\\
In a system of two coupled Van der Pol oscillators with a delay in the position coupling, we identify the amplitude death regions in the parameter space and conclude how does the presence of delay in the position coupling term influence amplitude death in the system.       
\section{Formulation}
Our model of coupled system of two identical Van der Pol oscillators is 
\begin{equation}
\ddot{x_1}+k\dot{x_1}(x_1^2-1)+\omega^2x_1+C(x_2(t-\tau)-x_1(t))=0
\end{equation}
\begin{equation}
\ddot{x_2}+k\dot{x_2}(x_2^2-1)+\omega^2x_2+C(x_1(t-\tau)-x_2(t))=0
\end{equation}
where $k>0$, $\omega$ is the natural frequency of each oscillator and the coefficient of position coupling term is $C>0$, $\tau$ is the delay which is a positive real number. As each system takes some finite time, $\tau$ to respond to the other one $\tau$ is positive real number. $\tau<0$ does not make any sense and as a result, the response of one oscillator depends on the past position of the other one $\tau$ time ago; that's why in the interaction term of each oscillator, the argument of position of the other one is $(t-\tau)$. The interaction term in this model is similar to the model taken by D. V. Ramana Reddy et al\cite{o} and J. Zhang et al\cite{q}. In the absence of any interaction, each individual oscillator has a stable limit cycle and one unstable fixed point at the origin. This can be confirmed by Li$\rm\hat{e}$nard's theorem\cite{g}. One can also analytically calculate time period and amplitude of such periodic oscillation in small $k$ and high $k$ limits by using approximation methods. In our model there is a fixed point at the origin $(x_1=x_2=\dot{x_1}=\dot{x_2}=0)$ and there may be a stable limit cycle\cite{t}\cite{u} surrounding it depending on the chosen values of parameters in the model. Our goal is to find out the amplitude death criteria that is the criteria of the change of stability of the fixed point. Keeping $k>0$, we look for Hopf bifurcation\cite{g}\cite{w} of the fixed point, i.e; the fixed point in the system becomes stable from unstable type and thus amplitude of the limit cycle surrounding it diminishes. This approach of finding amplitude death is also used by D. V. Ramana Reddy et al\cite{v}. The coupled system can be rewritten as
\begin{align}
&\dot{x_1}=y_1\\
&\dot{y_1}=-ky_1(x_1^2-1)-\omega^2 x_1-C(x_2(t-\tau)-x_1(t))\\
&\dot{x_2}=y_2\\
&\dot{y_2}=-ky_2(x_2^2-1)-\omega^2 x_2-C(x_1(t-\tau)-x_2(t))
\end{align}  
The above equations represent the behaviour of the coupled system in four dimensional phase space. The coordinate of the fixed point in this four dimensional space is $(x_1=0,~y_1=0,~x_2=0,~y_2=0)$. Around this coordinate writing the variations in $x_1,~y_1,~x_2$ and $y_2$ to be proportional to $e^{\lambda t}$, we get the condition for non trivial solution for the above variations  
\begin{equation}
\lambda^2-k\lambda+\omega^2-C=Ce^{-\lambda\tau}
\end{equation}
~~~~~~~~~~~~~~~~~~~~~~~~~~~~~~~~~~~~~~~~~~~~~~~~~~~Or,
\begin{equation}
\lambda^2-k\lambda+\omega^2-C=-Ce^{-\lambda\tau}\\
\end{equation}
We write
\begin{equation}
\lambda=\lambda_R+i\lambda_I
\end{equation} 
where $\lambda_R\in\mathbb{R}$ and $\lambda_I\in\mathbb{R}$. We first seek the conditions for Hopf bifurcation, i.e; the condition on $\lambda_R$ to change sign. Hence we put $\lambda_R$ to be zero in the above equation.
\begin{equation}
\lambda_I^2-\omega^2+C=Ccos(\lambda_I\tau)~{\rm and}~k\lambda_I=-Csin\lambda_I\tau
\end{equation} 
~~~~~~~~~~~~~~~~~~~~~~~~~~~~~~~~~~~~~~Or,
\begin{equation}
\lambda_I^2-\omega^2+C=-Ccos(\lambda_I\tau)~{\rm and}~k\lambda_I=Csin\lambda_I\tau
\end{equation}
The parameters defining the coupled system are $\omega,~k,~C$ and $\tau$. To eliminate $\lambda_I$ from the above equations, we first find the solutions for $\lambda_I$; given by
\begin{equation}
\lambda_{I1}=\sqrt{\frac{-(k^2+2C)+2\omega^2+\alpha(\omega,k,C)}{2}}
\end{equation}
\begin{equation}
\lambda_{I2}=\sqrt{\frac{-(k^2+2C)+2\omega^2-\alpha(\omega,k,C)}{2}}
\end{equation}
\begin{equation}
\lambda_{I3}=-\lambda_{I1}
\end{equation}
\begin{equation}
\lambda_{I4}=-\lambda_{I2}
\end{equation}
where
\begin{align*}
\alpha(\omega,k,C)=\sqrt{(k^2+2C)^2-4k^2\omega^2}
\end{align*}
Putting these values of $\lambda_I$ solutions, we get two families of curves for $\tau>0$.
\begin{align}
\tau_1(n,C)=\frac{(n+1)\pi-\theta}{\lambda_{I1}}~~{\rm and}~~\tau_2(n,C)=\frac{n\pi+\phi}{\lambda_{I2}}
\end{align}
where,
\begin{align}
&\theta=cos^{-1}\left(-\frac{k^2}{2C}+\frac{\alpha}{2C}\right),~\phi=cos^{-1}\left(\frac{k^2}{2C}+\frac{\alpha}{2C}\right)~{\rm and}~n\in\mathbb{N}~\cup~\lbrace0\rbrace
\end{align}
These two families of curves in the parameter space of the system represent the boundary curves along which the real part of $\lambda$ is zero and crossing these curves causes Hopf bifurcations, i.e; the stability of the limit cycle changes across these curves. 
\\There is another important point which is that the solutions which we find for $\lambda_{I}$ do not concern about $\lambda_I$ being real or not.\\
~~~~${\lambda_{Ii}\in \mathbb{R}}\Rightarrow\alpha\in\mathbb{R}\Rightarrow C\geq(\omega k-\frac{k^2}{2})$ where $i=1,~2,~3,~4$. Hence this value is the minimum value of $C,~C_{min}$.\\
{$C=C_{min},\lambda_{Ii}\in\mathbb{R}$}$\Rightarrow \omega>k$.\\ 
Now from the expressions, $\lambda_{I1,I3}^2>\lambda_{I2,I4}^2$.\footnote{As $\lambda_{I2}^2=\lambda_{I4}^2$ and $\lambda_{I1}^2=\lambda_{I3}^2$, we are using this kind of notation.}  Hence $\lambda_{I2,I4}^2>0\Rightarrow\lambda_{I1,I3}^2>0$. $\lambda_{I2,I4}^2>0\Rightarrow C\leq\frac{\omega^2}{2}$. Hence we have three restrictions on the family of curves.
\begin{align}
&(\omega k-\frac{k^2}{2})\leq C\leq \frac{\omega^2}{2}\\
&\omega\geq k
\end{align}
Thus for a given value of $\omega$ and $k$ such that $\omega\geq k$, the family of curves will be bounded within a window of $C$ which is $\Delta C$.
\begin{align}
\Delta C=\frac{1}{2}(\omega-k)^2
\end{align} 
Thus we find the boundary curves and the restrictions. Now we find the amplitude death regions, i.e; negative $\lambda_R$ regions in the parameter space. From equation (7), equation (8) and equation (9)
\begin{equation}
\lambda_I^2-\omega^2-\lambda_R^2+k\lambda_R+C=Ce^{-\lambda_R\tau}cos(\lambda_I\tau)
\end{equation} 
and
\begin{equation}
(k-2\lambda_R)\lambda_I=-Ce^{-\lambda_R\tau}sin\lambda_I\tau
\end{equation}
or,
\begin{equation}
\lambda_I^2-\omega^2-\lambda_R^2+k\lambda_R+C=-Ce^{-\lambda_R\tau}cos(\lambda_I\tau)
\end{equation} 
and
\begin{equation}
(k-2\lambda_R)\lambda_I=Ce^{-\lambda_R\tau}sin\lambda_I\tau
\end{equation} 
We calculate the quantities; $S+=\alpha k\lambda_{I1}^2\left(\frac{\delta\tau}{\delta\lambda_R}\right)\mid_{\lambda_{R}=0,~C=\rm constant}$ and $S-=\alpha k\lambda_{I1}^2\left(\frac{\delta\tau}{\delta\lambda_R}\right)\mid_{\lambda_{R}=0,~C=\rm constant}$. The sign of $S+$ and $S-$ give the changes of the signs of $\lambda_R$ across $\tau_1(n,C)$ family of curves and $\tau_2(n,C)$ family of curves respectively when the crossing is done along constant $C$ line for a given $\omega$ and $k$ with $\omega>k$. The positive coefficients $\alpha k\lambda_{I1}^2$ and $\alpha k\lambda_{I2}^2$ are there for the sake of calculation simplicity. $S+$ and $S-$ can be written as quadratic polynomials of $\tau$.
\begin{equation}
S+=A_0(\omega,~k,~C)+A_1(\omega,~k,~C)\tau+A_2(\omega,~k,~C)\tau^2
\end{equation}
where 
\begin{align*}
&A_0=2k\alpha-k(4(C-\omega^2)+k^2)\\
&A_1=\left(-k^2+\frac{\alpha}{2}\right)\alpha+\frac{k^2}{2}(k^2+4(C-\omega^2))-2C^2\\
&A_2=C^2k
\end{align*}
$A_0,~A_2>0$ for $C\in[C_{min}(=\omega k-\frac{k^2}{2}),C_{max}(=\frac{\omega^2}{2})]$.
\begin{equation}
S-=B_0(\omega,~k,~C)+B_1(\omega,~k,~C)\tau+B_2(\omega,~k,~C)\tau^2
\end{equation}
where
\begin{align*}
&B_0=2k\alpha+k(4(C-\omega^2)+k^2)\\
&B_1=\left(-k^2-\frac{\alpha}{2}\right)\alpha-\frac{k^2}{2}(k^2+4(C-\omega^2))+2C^2\\
&B_2=-C^2k
\end{align*}
It is not possible to find out the signs analytically. We do it numerically for $\omega=1,~k=0.1$ in the next section. 
\section{Results} 
We take $\omega$ to be 1 and $k$ to be 0.1 and we find $S+$ and $S-$ within the range of C. $C_{min}\simeq~0.1$, $C_{max}\simeq~0.5$ and $\Delta C\simeq~0.4$.
\begin{figure}[h!]
\includegraphics[scale=0.35]{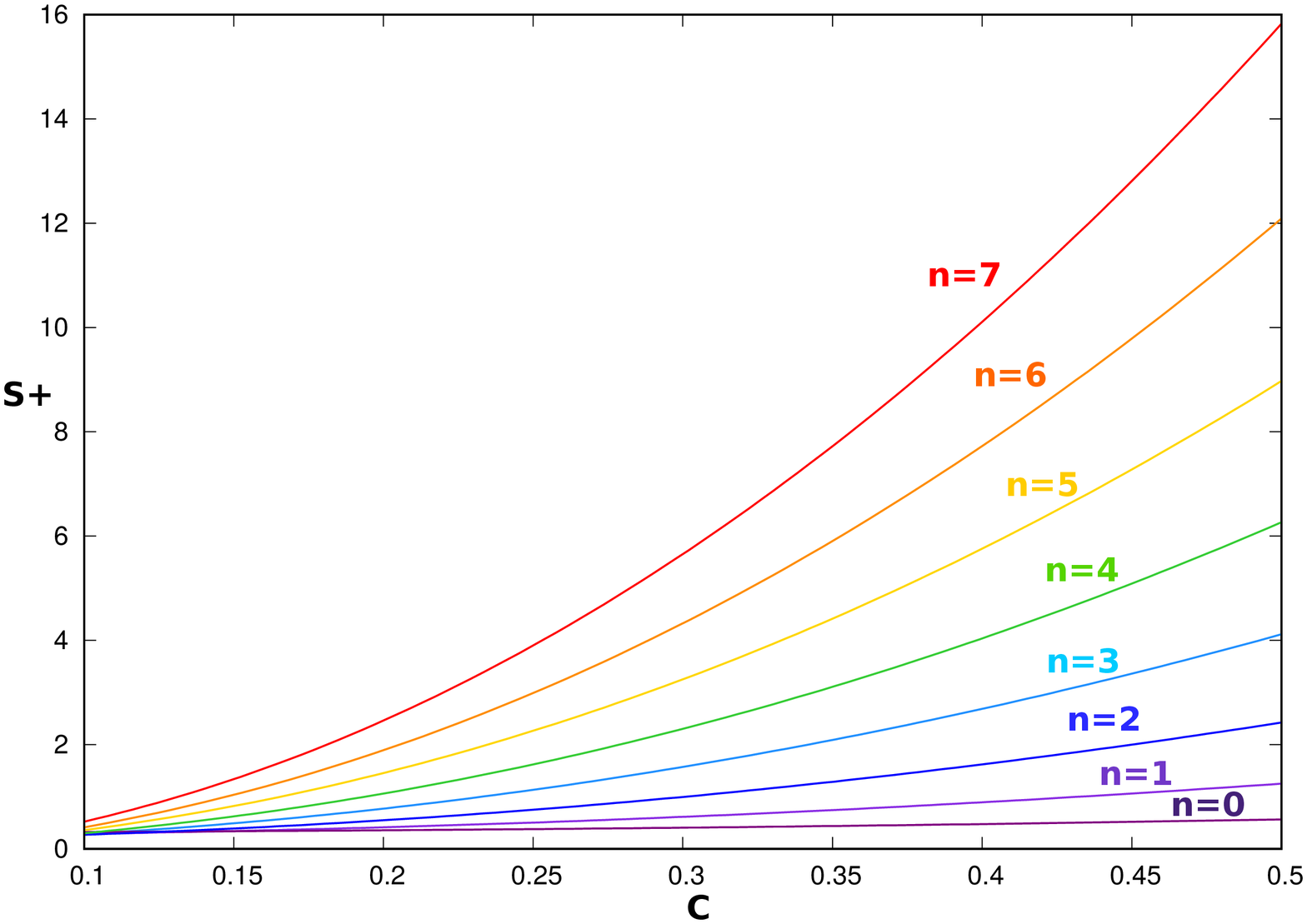}~~\includegraphics[scale=0.35]{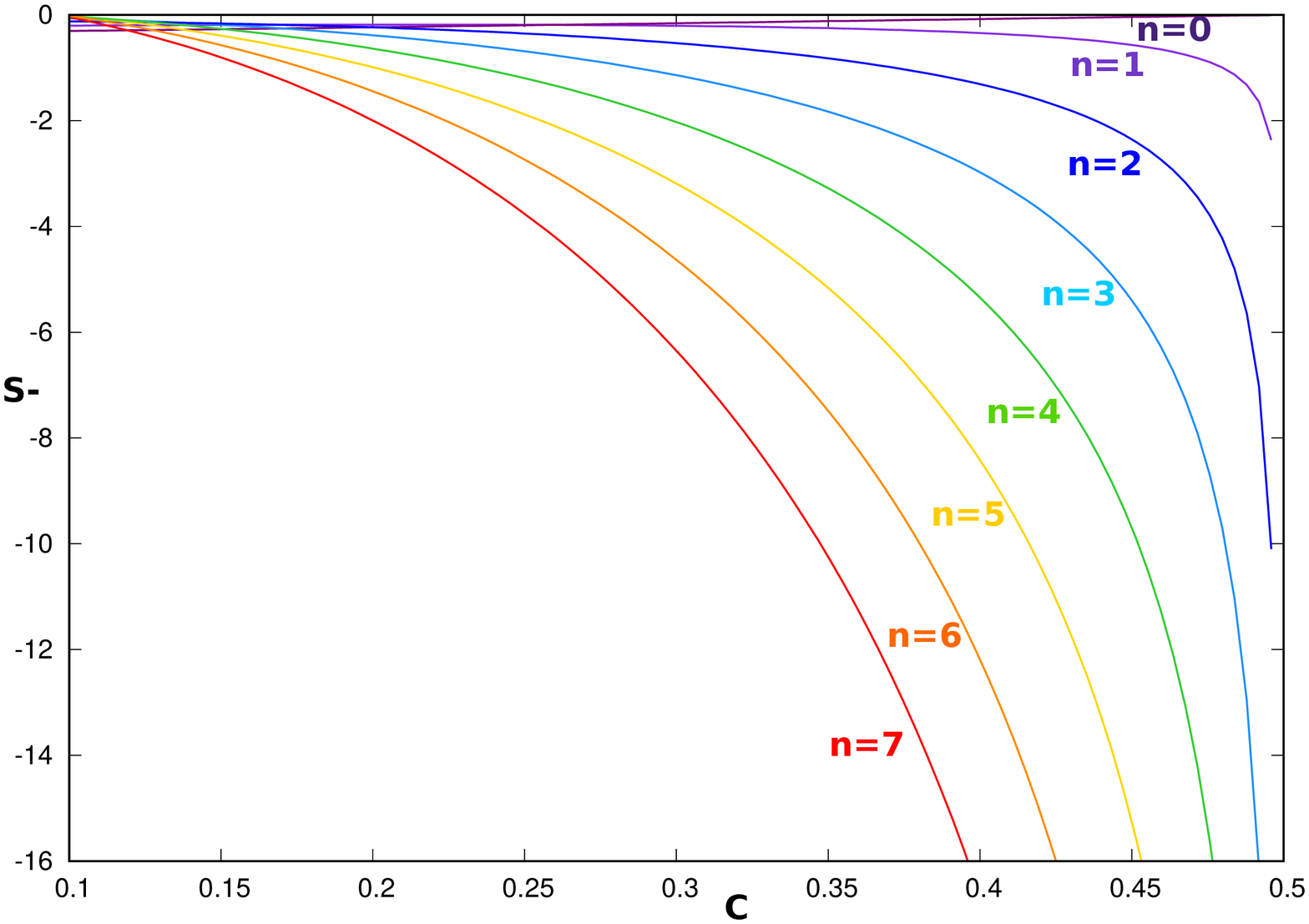}
\caption{$n$ in $\tau_1(n,C)$ and $\tau(n,C)$ is taken from $0$ to $7$. $S+$ is positive and $S-$ is negative within the range of C.}  
\end{figure}
\\
Hence $\lambda_R<0$ region is below the $\tau_1(n,C)$ curves and is above the $\tau_2(n,C)$ when viewed along constant $C$ lines while C axis is taken as x-axis.\\
The shaded regions in figure 2 are the intersections of these two regions such that $\lambda_R<0$.
The shaded regions in figure. 2 represent the amplitude death regions. Hence the shaded bounded regions in figure. 2 represents amplitude death islands. 
\begin{figure}[h!]
\centering
\includegraphics[scale=0.5]{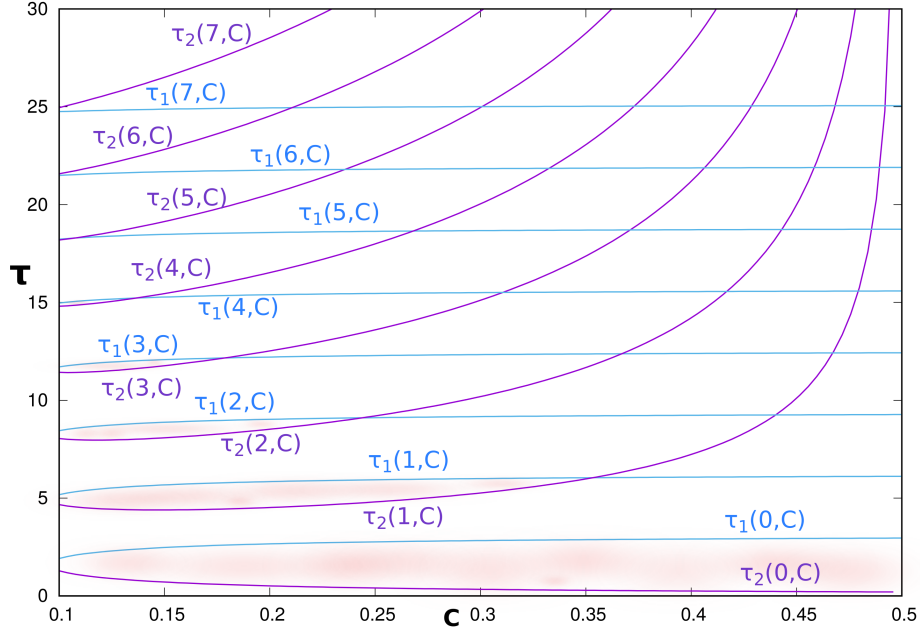} 
\caption{Five amplitude death islands are visible with decreasing area with the increase in order $n$.}
\end{figure}
\\
It seems from the figure that $\Delta C$ gives more or less the width of the amplitude death island of the lowest order,i.e; $n=0$.\\
In figure 2, the adjacent region corresponding to the lower half of $\tau_1(2,C)$ and the upper half of $\tau_2(1,C)$ does not represent amplitude death region because the real part of one pair of eigenvalue $\lambda$ has already become positive while crossing $\tau_1(1,C)$ along constant $C$ line in this region. Hence the real part of all the eigenvalues are not zero in this region. Same reasoning holds for the other regions of the same kind. 
\section{Summary and Conclusions}
We find that for $\omega>k$, the phenomena of amplitude death is possible to happen within a of finite window of $C$, $\Delta C$ which is proportional to the square of difference of $\omega$ and $k$. Amplitude death islands are found for $\omega=1$ and $k=0.1$. It is found that the width of the lowest order amplitude death island along $C-$ axis is roughly equal to $\Delta C~(=\frac{1}{2}(\omega-k)^2)$. One can also start with oscillators having same natural frequency, $\omega$; same coefficient of dissipation, $k$; same coefficient of coupling $C$ but different response time $\tau_1$ and $\tau_2$ ($\tau_1\neq\tau_2$); in that case, the whole analysis would be the same besides $\tau$ in the parameter space, has to be replaced by the average reaction time of the coupled system, i.e; $\tau_{avg}(=\frac{1}{2}(\tau_1+\tau_2))$.
\section{Acknowledgement}  
The author is thankful to Prof. Jayanta Kumar Bhattacharjee for useful discussions and suggestions.

\end{document}